\documentclass[12pt]{article}
\begin{document}

\title{Bigravity in Kucha\u{r}'s Hamiltonian formalism. 1. The general case
}
\author{Vladimir O. Soloviev and Margarita V. Tchichikina}
\maketitle
\begin{abstract}
  The Hamiltonian formalism of bigravity and massive gravity is studied here
for the general form of the interaction potential of two metrics. In the theories equipped 
with two spacetime metrics it is natural to use the Kucha\u{r} approach, because then
the role played by the lapse and shift variables becomes more transparent.
 We find conditions on the potential which are  necessary and sufficient for the existence of four first class constraints. The algebra of constraints is calculated
 in Dirac brackets formed on the base of all the second class constraints. It is
the celebrated algebra of hypersurface deformations. By fixing one 
metric we obtain a massive gravity theory free of first class constraints. Then we
can use symmetries of the background metric to derive conserved
quantities. These are ultralocal, if expressed in terms of the metric interaction potential. The special
case of potential providing less number of degrees of freedom is
treated in the companion paper.
\end{abstract}

\section{Introduction}
The idea to use more than one dynamical metric for the description of the real Universe  is not rather new. This theory is usually called multi-gravity. There is a hope that such a modification of the gravitational theory may allow to solve the dark energy problem. We restrict ourselves to the two metrics case, and will call such a model as bigravity, following pioneers~\cite{Kogan}. Suppose each metric interacts only with its own sort of matter, but metrics can interact with each other by means of some potential, i.e. a scalar density algebraically constructed of the two metrics. If we fix one metric by giving it the absolute meaning, we move from bigravity to a massive gravity (which can also be called a bimetric theory, as since N.~Rosen's~\cite{Rosen}  times the metric theories of gravity on the fixed background are called). In particular, massive gravity theories~\cite{DM} applied for explaining the dark energy phenomena in the Universe are in this category.

We consider the Hamiltonian formalism of bigravity in two papers dividing our presentation in two parts: the general case and the special case. In general case a concrete form of the interaction potential is not given, we only state that it is constructed algebraically (ultralocally, i.e. without derivatives), it is a scalar density and does not contain any fields of matter. We call as the special case the model with a potential that gives one degree of freedom less than the general case. The form of such a potential had been proposed recently~\cite{dRGT}. Unfortunately straightforward calculations with this potential in Hamiltonian formalism are difficult because of its matrix square root form, so, the analysis given already in a publications~\cite{HR} does not look transparent and decisive. We intend to derive the requirements on the potential which are necessary for realizing the program proclaimed in papers~\cite{dRGT} step-by-step, starting from a potential of the general form.
In contrast to the previous articles on this subject, we exploit not the Arnowitt-Deser-Misner (ADM)~\cite{ADM}, but the Kucha\u{r} approach~\cite{Kuchar} which preserves the freedom in spacetime coordinates choice and leads to the smaller number of spacetime metric components in the non-coordinate basis.

We denote as $f_{\mu\nu}$ and call as the first metric the metric that will be fixed (i.e. become a background one) after a shift from bigravity to massive gravity, and the second metric $g_{\mu\nu}$ which will always stay dynamical. Greek indices will run from $0$ to $3$, Latin -- from $1$ to $3$, signature is $(-,+,+,+)$.

Consider two copies of General Relativity Lagrangian, each copy supplied with its own sort of matter as a source (interaction with matter is supposed minimal), 
\begin{equation}
{\cal L}^{(f)}=\frac{1}{16\pi G^{(f)}}\sqrt{-f}f^{\mu\nu}R_{\mu\nu}^{(f)}+{\cal L}_M^{(f)}(\psi^A, f_{\mu\nu}),\label{eq:Lf}\
\end{equation}
\begin{equation}
{\cal L}^{(g)}=\frac{1}{16\pi G^{(g)}}\sqrt{-g}g^{\mu\nu}R_{\mu\nu}^{(g)}+{\cal L}_M^{(g)}(\phi^A, g_{\mu\nu}),\label{eq:Lg}
\end{equation}
where $f_{\mu\nu}$ and $g_{\mu\nu}$ are the first and the second spacetime metrics correspondingly, $f$ and $g$ are their determinants, $R_{\mu\nu}^{(f)}$ and $R_{\mu\nu}^{(g)}$ are Ricci tensors, $G^{(f)}$ and $G^{(g)}$ are gravitational constants, ${\cal L}_M^{(g)}$ and ${\cal L}_M^{(f)}$ are Lagrangians of the first and the second matter, $A$ denotes abstract indicies for matter fields, and construct of them a new Lagrangian by substracting the potential
\begin{equation}
\sqrt{-f}U( f_{\mu\nu},g_{\mu\nu}),
\end{equation}  
which has the dimension of the fourth power of mass. Hence,
\begin{equation}
{\cal L}={\cal L}^{(f)}+{\cal L}^{(g)}-\sqrt{-f}U( f_{\mu\nu}, g_{\mu\nu}).\label{eq:L_bi}
\end{equation}
It is clear that a choice of metric  determinant  which is present explicitly is not of importance because function $U( f_{\mu\nu},g_{\mu\nu})$ may arbitrarily depend on the ratio of two determinants.

\section{The Kucha\u{r} Hamiltonian formalism in General Relativity}
Let the action of General Relativity (GR) is written as follows
\begin{equation}
S=\int{\cal L} d^4X, 
\end{equation}
where $X^\alpha$ are spacetime coordinates, and the form of Lagrangian is the same as in formulas (\ref{eq:Lf}), (\ref{eq:Lg}). 

In the construction of Hamiltonian formalism it is necessary to separate explicitly the time coordinate from the spatial ones. The state should be determined by values of the gravitational and matter fields given at all points of space at one definite moment of time. So, the state is to be given on a spacelike hypersurface embedded in  spacetime. The flow of time corresponds to the motion of this hypersurface through spacetime or to the continuous transform from one hypersurface to another, so we need one-parametrical family of spacelike hypersurfaces. The role of time coordinate $t$ is to be played by a parameter which continuously and monotonically numerates hypersurfaces.  It is suitable to introduce spatial coordinates  $x^i$ on the initial hypersurface and then continue them to all hypersurfaces in such a way that lines going through the points with the same values of coordinates can be treated as observer world lines, i.e. they should be timelike. 

In ADM approach~\cite{ADM}, which is the most popular, the choice of spacelike hypersurfaces family is determined by the choice of spacetime coordinate frame:
\begin{equation}
t=X^0,\quad x^i=X^i,\quad \gamma_{ij}=g_{ij},\quad  N=(-g^{00})^{-1/2},\quad N_i=g_{0i}.
\end{equation}
There is another approach proposed by Kucha\u{r}~\cite{Kuchar} where two coordinate systems are in action, one  is an arbitrary spacetime coordinate system $X^\alpha$, the other  $(t,x^i)$ is defined by embedding variables 
\begin{equation}
X^{\alpha}=e^{\alpha}(x^i, t),
\end{equation}
where time is a parameter monotonically numerating hypersurfaces and other three coordinates numerate points on a hypersurface. 
We will follow this fully covariant Kucha\u{r}'s method.
Then fields
\begin{equation}
e^\alpha_i\equiv\frac{\partial e^{\alpha}}{\partial x^i}
\end{equation}
will simultaneously be vectors in spacetime and covectors in space. The metric induced on a hypersurface is given as follows
\begin{equation}
 \gamma_{ij}=g_{\mu\nu}e^\mu_i e^\nu_j.
\end{equation}
As usual, inverse matrices to metrics $g_{\mu\nu}$ and $\gamma_{ij}$ are denoted as $g^{\mu\nu}$ and $\gamma^{ij}$. Moving indices (both Greek and Latin) up and down is provided by means of them:
\begin{equation}
{\bar e}_{\alpha i}=g_{\alpha\beta}e^\beta_i,\quad {\bar e}^i_{\alpha}= g_{\alpha\beta}e^\beta_j\gamma^{ij}.
\end{equation}
We will  use here a bar to make a distinction between the quantities defined by means of metric and the quantities $e^\alpha_i$ which are independent of the metric. In the following, when we will work with two metrics, the bar will stay connected with    $g_{\mu\nu}$.

Next we introduce normal 1-form
\begin{equation}
n_\alpha e^{\alpha}_i=0,
\end{equation}
which can be converted with the help of metric tensor to a normalized vector:
\begin{equation}
{\bar n}^\alpha=g^{\alpha\beta}{\bar n}_\beta\label{eq:unitnormal}, \qquad g^{\mu\nu}{\bar n}_\mu {\bar n}_\nu=-1.
\end{equation}
We can use basis $({\bar n}^\alpha, e^\alpha_i)$ to decompose every vector or tensor in spacetime, for example, Ricci tensor (with account for its symmetric nature):
\begin{equation}
R^{\mu\nu}=R^{\perp\perp}{\bar n}^{\mu}{\bar n}^{\nu}+
R^{i\perp}(e^{\mu}_{i}{\bar n}^{\nu}+ {\bar n}^{\mu}e^{\nu}_{i})
+R^{ij}e^{\mu}_{i}e^{\nu}_{j},
\end{equation}
here the components are calculated as follows:
\begin{equation}
R^{\perp\perp}=R^{\mu\nu}{\bar n}_\mu {\bar n}_\nu, \qquad R^{i\perp}=-R^{\mu\nu}{\bar n}_\mu {\bar e}_{\nu i}, \qquad R^{ij}=R^{\mu\nu}{\bar e}_{\mu i}{\bar e}_{\nu j}.
\end{equation}

With the known technique~\cite{Kuchar,Solo88}, based on expression of the Riemann tensor projections as the corresponding projections of covariant derivatives commutators, it is possible to write GR Lagrangian density containing the single spacetime metric
 $g_{\mu\nu}$ in the following form
\begin{equation}
{\cal L}^{(g)}=
\frac{{\bar N}\sqrt{\gamma}}
{\kappa^{(g)}}
\left(R^{(\gamma)}-{\bar K}^2+Sp {\bar K}^2 \right)+{\cal L}_M^{(g)}(\phi^A,\gamma_{ij},{\bar N},{\bar N}^i),\label{eq:short-L}
\end{equation}
where boundary terms, i.e. total time derivatives and spatial divergences are ignored, as we  do not discuss boundary conditions. Here $\kappa^{(g)}=1/16\pi G^{(g)}$, $R^{(\gamma)}$ is  the scalar curvature of the induced metric, $\gamma=\mathrm{det}||\gamma_{ij}||$, lapse and shift functions ${\bar N}, {\bar N}^i$ are components of the decomposition of time vector field over the basis constructed with metric $g_{\mu\nu}$ 
\begin{equation}
N^\alpha\equiv \frac{\partial X^\alpha}{\partial t}={\bar N}{\bar n}^\alpha+{\bar N}^i e^\alpha_i.
\end{equation}
The second fundamental form ${\bar K}_{ij}$ arising in the Lagrangian density can be expressed through the already introduced variables by formula: 
\begin{equation}
{\bar K}_{ij}=\frac{1}{2{\bar N}}\left({\bar N}_{i|j}+{\bar N}_{j|i}-\dot\gamma_{ij}\right),
\end{equation}
here ${\bar K}=\gamma^{ij} {\bar K}_{ij}$, $\mathrm{Sp} {\bar K}^2={\bar K}_{ij}{\bar K}^{ij}$,
vertical bar denotes a covariant derivative defined by induced metric.   The momenta conjugate to variables $\gamma_{ij}$ and $\phi^A$ are determined as follows:
\begin{equation}
\pi^{ij}=\frac{\partial{\cal L}^{(g)}}{\partial\dot\gamma_{ij}}=
-\frac{\sqrt{\gamma}}{\kappa^{(g)}}
({\bar K}^{ij}-\gamma^{ij}
 {\bar K}),\quad \pi_A=\frac{\partial{\cal L}}{\partial\dot\phi^A},
\end{equation}
and the momenta conjugate to $\bar N$ and ${\bar N}^i$ are zero, as expression (\ref{eq:short-L}) does not contain velocities $\dot {\bar N}$, $\dot {\bar N}^i$
\begin{equation}
\pi_{\bar N}=0,\qquad \pi_{{\bar N}^i}=0,\label{eq:primary}
\end{equation}
so, these equations are primary constraints.
The velocities for gravitational variables $\gamma_{ij}$ can be expressed by means of momenta
\begin{equation}
\dot{\gamma_{ij}}= {\bar N}_{i|j}+ {\bar N}_{j|i}+\frac{2\kappa^{(g)} {\bar N}}{\sqrt{\gamma}}
(\pi_{ij}-\gamma_{ij}\frac{\pi}{2}).
\end{equation}
We believe that it is possible to do the same for the matter variables (in other case we can analyse new constraints arising there), and after performing the Legendre transform 
\begin{equation}
{\rm H}_{\rm canonical}=\int d^3x\left(
\pi^{ij}\dot\gamma_{ij}+\pi_A\dot\phi^A-{\cal L} 
\right),
\end{equation}
to obtain the canonical Hamiltonian 
 (up to surface terms)  
\begin{equation}
{\rm H}_{\rm canonical}=\int d^3x\left(
{\bar N}{\bar{\cal H}}+{\bar N}^i{\bar{\cal H}}_i 
\right).\label{eq:Ham_can}
\end{equation}
The Poisson brackets for functionals of canonical variables are as follows
\begin{equation}
\{F,G\}=\int d^3x\left(\frac{\delta F}{\delta {\bar N}}\frac{\delta G}{\delta \pi_{\bar N}}+\frac{\delta F}{\delta {\bar N}^i}\frac{\delta G}{\delta \pi_{{\bar N}^i}}+
\frac{\delta F}{\delta \gamma_{ij}}\frac{\delta G}{\delta\pi^{ij}}+\frac{\delta F}{\delta \phi^{A}}\frac{\delta G}{\delta\pi_{A}} - (F\leftrightarrow G)\right).\label{eq:PB_GR}
\end{equation}
According to the Dirac method~\cite{Dirac}, one should further go to the extended Hamiltonian by adding the primary constraints with arbitrary Lagrangian multipliers
\begin{equation}
{\rm H}_{\rm extended}=\int d^3x\left(
{\bar N}{\bar{\cal H}}+{\bar N}^i{\bar{\cal H}}_i 
+\lambda\pi_{\bar N}+\lambda^i\pi_{{\bar N}^i}
\right).\label{eq:Ham_ext} 
\end{equation}
Then expressions 
\begin{equation}
{\bar{\cal  H}}=-\frac{1}{\sqrt{\gamma}}\left(\frac{1}{\kappa^{(g)}}\gamma R^{(\gamma)}+\kappa^{(g)}\left(\frac{\pi^2}{2}-\mathrm{Sp}\pi^2 \right) \right)+{\bar{\cal H}}_M\label{eq:constraint1}
\end{equation}
 and
 \begin{equation}   
  {\bar  {\cal  H}}_i=-2\pi_{i|j}^j+{\bar{\cal H}}_{iM}\label{eq:constraint2}
 \end{equation}   
     become secondary constraints, as their equality to zero is necessary for primary constraints staying zero in evolution. Here $\pi=\gamma_{ij}\pi^{ij}$, $\mathrm{Sp}\pi^2=\pi^{ij}\pi_{ij}$, ${\bar{\cal H}}_M, {\bar{\cal H}}_{iM}$ is the matter contribution.
      
The standard procedure requires to remove four pairs of canonical variables
 ${\bar N},{\bar N}^i$, $\pi_{\bar N},\pi_{{\bar N}^i}$ and of primary constraints  (\ref{eq:primary}) с by putting on four gauge conditions which really  change only the meaning of four letters, i.e.  replace canonical variables with functions (we do not like to introduce new notations and we make this change  only here)
\begin{equation}
{\bar N}-{\bar {\cal N}}(x)=0,\qquad {\bar N}^i-{\bar {\cal N}}^i(x)=0. 
\end{equation}
These gauges have nonzero Poisson brackets  with primary constraints  (\ref{eq:primary}) and allow to treat all of them together as eight second class constraints, this leads to Dirac brackets:
\begin{equation}
\{F,G\}_D=\int d^3x\left(\frac{\delta F}{\delta \gamma_{ij}}\frac{\delta G}{\delta\pi^{ij}}+\frac{\delta F}{\delta \phi_{A}}\frac{\delta G}{\delta\pi^{A}} - (F\leftrightarrow G)\right).\label{eq:DB_GR}
\end{equation}
The number of gravitational degrees of freedom is now determined by a simple calculation: take the number of induced metric $\gamma_{ij}$ independent  components  and subtract the number of first class constraints: $6-4=2$, now Hamiltonian takes a following form 
\begin{equation}
{\rm H}_{\rm partially\ reduced}=\int d^3x\left(
{\bar {\cal N}}{\bar{\cal H}}+{\bar {\cal N}}^i{\bar{\cal H}}_i 
\right).\label{eq:Ham1}
\end{equation}
Avoiding cumbersome terminology and notations we as usual will call Dirac brackets   (\ref{eq:DB_GR}) as Poisson brackets,  the partially reduced Hamiltonian (\ref{eq:Ham1}) as  Hamiltonian, ${\bar N}={\bar {\cal N}}$, ${\bar N}^i={\bar {\cal N}}^i$  as Lagrangian multipliers standing before constraints (\ref{eq:constraint1}), (\ref{eq:constraint2}).  These constraints are first class as their algebra is as follows 
\begin{equation}
 \{ {\bar{\cal H}}(x),{\bar{\cal H}}(y)\}=
 (
 \gamma^{ik}(x){\bar{\cal H}}_k(x)+ \gamma^{ik}(y){\bar{\cal H}}_k(y)
 )
 \delta_{,i}(x,y),\label{eq:alg1}
\end{equation}
\begin{equation}
  \{ {\bar{\cal H}}_i(x),{\bar{\cal H}}_k(y)\}={\bar{\cal H}}_i(y)\delta_{,k}(x,y)+ {\bar{\cal H}}_k(x)\delta_{,i}(x,y),\label{eq:alg2}
\end{equation}
\begin{equation}
  \{ {\bar{\cal H}}_i(x),{\bar{\cal H}}(y)\}={\bar{\cal H}}(x)\delta_{,i}(x,y),\label{eq:alg3}
\end{equation}
 reflecting a freedom of hypersurface deformations in Riemannian space. That we have four first class constraints with arbitrary Lagrangian multipliers implies the possibility of imposing four gauge conditions, and the gravitational field in GR, described by variables $\gamma_{ij}$, therefore has $6-4=2$ degrees of freedom.

\section{The Hamiltonian approach to bigravity}
Having two spacetime metrics  we get  two induced metrics on a spatial hypersurface:
\begin{equation}
\gamma_{ij}=g_{\mu\nu}e^\mu_i e^\nu_j,\qquad \eta_{ij}=f_{\mu\nu}e^\mu_i e^\nu_j,
\end{equation}
and also two different unit normal vectors that will be denoted as $n^\alpha$ and ${\bar n}^\alpha$:
\begin{equation}
n_\alpha e^\alpha_i=0={\bar n}_\alpha  e^\alpha_i, \quad f^{\alpha\beta}n_\alpha n_\beta=-1, \quad g^{\alpha\beta} {\bar n}_\alpha{\bar n}_\beta=-1,
\end{equation}
and two bases $(n^\alpha,e^\alpha_i)$,  $({\bar n}^\alpha,e^\alpha_i)$. Moving indices up and down is provided by the corresponding metric tensors $f_{\mu\nu},\eta_{ij}$ and $g_{\mu\nu},\gamma_{ij}$.
We will decompose any spacetime vectors and tensors over basis $(n^\alpha, e^\alpha_i)$, constructed with the help of $f_{\mu\nu}$ which will be called the first metric. For example, we decompose the second metric tensor (with account for its symmetry)   as follows
\begin{equation}
g^{\mu\nu}=g^{\perp\perp}n^{\mu}n^{\nu}+
g^{\perp i}(e^{\mu}_{i}n^{\nu}+ n^{\mu}e^{\nu}_{i})
+g^{ij}e^{\mu}_{i}e^{\nu}_{j},\label{eq:g}
\end{equation}
whereas,
\begin{equation}
f^{\mu\nu}=-n^\mu n^\nu +e^\mu_i e^\nu_j \eta^{ij},\label{eq:f}
\end{equation}
the components are calculated by formulas:
\begin{equation}
g^{\perp\perp}=g^{\mu\nu}n_\mu n_\nu, \quad g^{\perp i}=-g^{\mu\nu}n_\mu e_\nu^i, \quad g^{ij}=g^{\mu\nu}e_\mu^ie_\nu^j=\gamma^{ij}+\frac{g^{\perp i}g^{\perp j}}{g^{\perp\perp}}.\label{eq:project}
\end{equation}
We write each metric contribution to bigravity Lagrangian (\ref{eq:L_bi}) in the similar way to equation (\ref{eq:short-L}).
 The potential, in its turn, depends on the two spacetime metrics, therefore it is necessary to brake their symmetry and take one of the metrics for construction of the basis in order to transform the potential to $(3+1)$-form.   Previously we already have decided to call $f_{\mu\nu}$ the first metric and to use the basis constructed with its help. Then besides lapse $N$  and shift $N^i$ 
  \begin{equation}
  N^\alpha=Nn^\alpha+N^i e^\alpha_i={\bar N}{\bar n}^\alpha+{\bar N}^i e^\alpha_i
  \end{equation}
and induced spatial metrics $\eta_{ij}$, $\gamma_{ij}$, the potential  depends
on other four components of metric  $g^{\mu\nu}$ decomposition over basis $(n^\alpha, e^\alpha_i)$, let us accentuate that in contrast to $N$, quantities $g^{\perp\perp}, g^{\perp i}$ enter the potential in a nonlinear way. Relations between the two bases
\begin{equation}
{\bar n}^\alpha=\sqrt{-g^{\perp\perp}}n^\alpha-
\frac{g^{\perp i}}{\sqrt{-g^{\perp\perp}}}e^\alpha_i
\end{equation}
allow us to express lapse and shift of metric $g_{\mu\nu}$, i.e. ${\bar N}$, ${\bar N^i}$ through $N$, $N^i$ and normal projections   $g^{\perp\perp}, g^{\perp i}$:
\begin{equation}
 {\bar N}=\frac{ N}{\sqrt{-g^{\perp\perp}}},\qquad {\bar N^i}= N^i-\frac{g^{\perp i}}{g^{\perp\perp}}N.\label{eq:NN}
\end{equation}
It is suitable for the following to introduce new variables
\begin{equation}
u=\frac{1}{\sqrt{-g^{\perp\perp}}},\qquad u^i=-\frac{g^{\perp i}}{g^{\perp\perp}},
\end{equation}
 having the simple geometric meaning: $u$ is an inverse of a norm (calculated in the second metric) of vector $n^\alpha$, constructed as a unit normal (in the first metric) to the hypersurface, and $u^i$ are three projections (calculated in the second metric) of coordinate basis vectors onto this unit normal
\begin{equation}
u=\frac{1}{\sqrt{|g^{\mu\nu}n_\mu n_\nu|}}, \qquad u^i=\frac{g^{\mu\nu}n_\mu e_\nu^i}{\sqrt{|g^{\mu\nu}n_\mu n_\nu|}}.
\end{equation}
Then equations (\ref{eq:NN}) takes a form
\begin{equation}
 {\bar N}=u N, \qquad {\bar N^i}= N^i +u^i N.
\end{equation}
We consider as dynamic variables two sets of matter fields $\phi_A, \psi_A$, two induced metrics on the hypersurface $\eta_{ij},\gamma_{ij}$, components of the time vector in the chosen basis  (i.e. lapse and shift) $ N,  N^i$ and four additional variables $u,u^i$, taken intead of the second metric projections $g^{\perp\perp}, g^{\perp i}$. 
Momenta are defined in the usual way
\begin{equation}
\Pi^{ij}=\frac{\partial{\cal L}}{\partial\dot\eta_{ij}},\quad \pi^{ij}=\frac{\partial{\cal L}}{\partial\dot\gamma_{ij}},\quad \Pi_A=\frac{\partial{\cal L}}{\partial\dot\psi^A},\quad  \pi_A=\frac{\partial{\cal L}}{\partial\dot\phi^A}.
\end{equation}
The two procedures of defining momenta and Legendre transform are realized in parallel and independent way for both terms ${\cal L}^{(f)}$ and ${\cal L}^{(g)}$. 
Primary constraints arise because velocities of eight variables $N,N^i,u,u^i$ are absent in the Lagrangian density,  these constraints have a following form
\begin{equation}
\pi_{N}=0,\quad \pi_{N^i}=0,\label{eq:piN}
\end{equation} 
\begin{equation}
\pi_{u}=0, \quad \pi_{u^i}=0.\label{eq:piu}
\end{equation}
Certainly, some additional primary constraints can appear due to gauge invariance which matter fields may have, but we will not discuss this here. The canonical Hamiltonian is derived by the Legendre transform
\begin{equation}
{\rm H}_{\rm canonical}=\int d^3x \left( \Pi^{ij}\dot{\eta}_{ij}+\pi^{ij}\dot{\gamma}_{ij}+\pi_{\psi^A}\dot{\psi^A}+\pi_{\phi^A}\dot{\phi^A}-{\cal L} \right)
\end{equation}
and expression of velocities through the corresponding momenta, this provides us with the two similar looking terms without common variables plus the potential: 
\begin{equation}
{\rm H}_{\rm canonical}=\int d^3x\left(N{\cal H}+N^i{\cal H}_i+{\bar N}{{\bar{\cal H}}}+\bar{N^i}{\bar{\cal H}}_i+N\sqrt{\eta}U\right),
\end{equation}
or in other way,
\begin{equation}
{\rm H}_{\rm canonical}=\int d^3x \left(N\left({\cal H}+u{\bar{\cal H}}+u^i{\bar{\cal H}}_i+\sqrt{\eta}U \right)+ N^i\left({{\cal H}}_i+\bar{\cal H}_i \right)\right).\label{eq:H}
\end{equation}
The canonical Poisson brackets have a standard appearance corresponding to the set of conjugate variables $(\eta_{ij},\Pi^{ij})$, $(\gamma_{ij},\pi^{ij})$, $(\psi^A,\Pi_A)$, $(\phi^A,\pi_A)$, $(N,\pi_N)$, $(N^i,\pi_{N^i})$, $(u,\pi_u)$, $(u^i,\pi_{u^i})$. 
The requirement of conservation of the primary constraints  (\ref{eq:piN}), (\ref{eq:piu}) in Hamiltonian evolution forces us to put their Poisson brackets with Hamiltonian to zero 
\begin{equation}
\{\pi_N,{\rm H}\}=0, \quad \{\pi_{N^i},{\rm H}\}=0, \quad \{\pi_u,{\rm H}\}=0, \quad \{\pi_{u^i},{\rm H}\}=0,
\end{equation} 
this leads to the secondary constraints:
\begin{equation}
{\cal R}\equiv{{\cal H}}+u{\bar{\cal H}}+u^i{\bar{\cal H}}_i+\tilde U=0,\label{eq:R}
\end{equation}
\begin{equation}
{\cal R}_i\equiv{{\cal H}}_i+\bar{\cal H}_i=0,\label{eq:Ri}
\end{equation}
\begin{equation}
{\bar{\cal H}}+\frac{\partial\tilde{U}}{\partial u}=0,\qquad {\bar{\cal H}}_i+\frac{\partial\tilde{U}}{\partial u^i}=0,\label{eq:last}
\end{equation}
where $\tilde{U}=\sqrt{\eta}U$.
In the general case, which is a subject of this paper, the four last constraints together with the four last primary constraints occurs second class~\cite{BD} and can be excluded from Hamiltonian after introduction of Dirac brackets (see Appendix \ref{S:DB}). The special case is treated in the companion article~\cite{SolTchi2}.  For other secondary constraints the algebra of hypersurface deformations should be satisfied in the Dirac brackets: 
\begin{equation}
 \{ {\cal R}(x),{\cal R}(y)\}_D=\left[\eta^{ik}(x){\cal R}_k(x)+ \eta^{ik}(y){\cal R}_k(y)\right]\delta_{,i}(x,y),\label{eq:bialg1}
\end{equation}
\begin{equation}
  \{ {\cal R}_i(x),{\cal R}_k(y)\}_D={\cal R}_i(y)\delta_{,k}(x,y)+ {\cal R}_k(x)\delta_{,i}(x,y),\label{eq:bialg2}
\end{equation}
\begin{equation}
  \{ {\cal R}_i(x),{\cal R}(y)\}_D={\cal R}(x)\delta_{,i}(x,y).\label{eq:bialg3}
\end{equation}
Calculating l.h.s. first as Poisson brackets, then applying second class constraints   (\ref{eq:last}), at last  comparing results with r.hs., we derive conditions on the potential given as function of variables $\eta_{ij},\gamma_{ij},u,u^i$ (details see in Appendix \ref{S:FC}):
\begin{equation}
 2\eta_{jk}\frac{\partial\tilde{U}}{\partial\eta_{ij}}+2\gamma_{jk}\frac{\partial\tilde{U}}{\partial\gamma_{ij}}-u^i\frac{\partial\tilde{U}}{\partial u^k}=\delta^i_k\tilde{U},\label{eq:pot1}
\end{equation}
\begin{equation}
 2u^j\gamma_{jk}\frac{\partial\tilde{U}}{\partial\gamma_{k\ell}}-u^\ell u\frac{\partial\tilde{U}}{\partial u}+\left(\eta^{k\ell}-u^2\gamma^{k\ell}-u^k u^\ell \right)\frac{\partial\tilde{U}}{\partial u^k}=0.\label{eq:pot2}
\end{equation}
These conditions preserve their form under a change of variables corresponding to the interchange of bases and roles of the two metrics:
\begin{equation}
u\rightarrow\frac{1}{u},\quad u^i\rightarrow -\frac{u^i}{u},\quad \eta_{ij}\leftrightarrow\gamma_{ij},\quad \tilde U\rightarrow \frac{\tilde U}{u}. 
\end{equation}
The number of gravitational degrees of freedom $N_{\rm DOF}$ in the bigravity theory with the general potential is equal to $N_{\rm DOF}=n-m-\ell/2$, where $n$ is the number of pairs of canonically conjugate gravitational variables, $m$ is the number of first class constraints, and $\ell$ is the number of second class constraints. Here, such canonically conjugate variables are $(\eta_{ij},\Pi^{ij})$, $(\gamma_{ij},\pi^{ij})$, $(u,\pi_u)$, and $(u^i,\pi_{u^i})$, the first class constraints are ${\cal R}$, ${\cal R}_i$, and the second class constraints are $\pi_u$, $\pi_{u^i}$, ${\cal S}$, and ${\cal S}_i$. We therefore have $n=16$, $m=4$, and $\ell=8$. As the result, we obtain eight gravitational degrees of freedom, which correspondes to one massless and one massive gravitational field. The massive field then has six degrees of freedom , which indicates the presence of an extra nonphysical component, a ghost. The sign of the expression for the Hamiltonian density is indefinite, which, generally speaking, may indicate an instability.  

The simplest method to derive Hamiltonian equations is to use canonical Hamiltonian and Poisson brackets. Then they are looking similar to GR Hamiltonian equations (here nabla denotes a covariant derivative defined by $\eta_{ij}$):
\begin{eqnarray}
\dot{\eta}_{ij}&=&\{\eta_{ij},{\rm H}^{f}\}=  \nabla_j N_{i}+  \nabla_i N_{j}+\kappa^{(f)}\frac{2 {N}}{\sqrt{\eta}}
(\Pi_{ij}-\eta_{ij}\frac{\Pi}{2}),\\
\dot{\gamma}_{ij}&=&\{\gamma_{ij},{\rm H}^{g}\}=  {\bar N}_{i|j}+  {\bar N}_{j|i}+\kappa^{(g)}\frac{2 {\bar N}}{\sqrt{\gamma}}
(\pi_{ij}-\gamma_{ij}\frac{\pi}{2}),\\
\dot{\Pi}^{ij}&=&\{\Pi^{ij},{\rm H}^{f}\}-N\frac{\partial {\tilde U}}{\partial\eta_{ij}},\\
\dot{\pi}^{ij}&=&\{\pi^{ij},{\rm H}^{g}\}-N\frac{\partial {\tilde U}}{\partial\gamma_{ij}},\\
\dot\psi^A&=&\{\psi^A,H^{(f)}\},\\
\cdots & = & \cdots
\end{eqnarray}
but one should remember that if
 $N,N^i$ are arbitrary Lagrangian multipliers (before putting on gauge conditions),  $u,u^i$   after solving second class constraints become functions of  $\bar{\cal H}, \bar{\cal H}_i$,  which in their turn are expressed by equations (\ref{eq:constraint1}), (\ref{eq:constraint2}) through canonical variables. Certainly, these Hamiltonian equations can be derived by means of Dirac brackets from Hamiltonian after applying second class constraints to it:
\begin{equation}
{\rm H}=\int d^3x\left( N\left({\cal H}+\tilde U -u\frac{\partial\tilde U}{\partial u}-u^i\frac{\partial\tilde U}{\partial u^i} \right)+N^i\left({\cal H}_i-\frac{\partial \tilde U}{\partial u^i}\right)\right).\label{eq:66}
\end{equation}
When we are given embedding variables $X^\alpha= e^\alpha(t,x^i)$, we can express both spacetime metrics  in coordinates  $X^\alpha$, if we first find the unit normal vector
\begin{equation}
n^\alpha=\frac{1}{N}\left(\frac{\partial e^\alpha}{\partial t}-N^ie^\alpha_i\right),
\end{equation}
and then apply formulas (\ref{eq:g}), (\ref{eq:f}), (\ref{eq:project}) for two metrics decompositions in this basis.

\section{Hamiltonian approach in massive gravity}
Consider now the situation when only the second metric is dynamical and the first one is  background, i.e. it is a fixed solution of GR equations. Traditionally~\cite{Rosen} such theories are called bimetric. Recently they attracted a lot of attention and some valuable reviews have appeared~\cite{Rubak,Blas,MMMM,Hinter}.

If metric tensor $f_{\mu\nu}$ and embedding variables $e^\alpha(t,x^k)$ are given, then  functions
 $N(t,x^k)$, $N^i(t,x^k)$, $\eta_{ij}(t,x^k)$ are determined. We can take expression (\ref{eq:H}), derived above for bigravity,  as a Hamiltonian, treating  $N,N^i$ not as variables, but as given parameters, and taking into account that ${\cal H},{\cal H}_i$ are now zero,
\begin{equation}
H=\int d^3x \left(N\left(u{\bar{\cal H}}+u^i{\bar{\cal H}}_i+\tilde U \right)
+ N^i\bar{\cal H}_i \right).
\end{equation}
Then primary constraints are only equations (\ref{eq:piu}), and they lead to secondary constraints (\ref{eq:last}). In the general case (remember, that special case is treated in companion paper~\cite{SolTchi2}) all these constraints are second class, and Dirac brackets constructed from them will coincide with Poisson brackets for the functionals which depend only on the independent variables $\gamma_{ij}$, $\pi^{ij}$, $\phi^A$, $\pi_A$:  
\begin{equation}
\{F,G\}_D=\int d^3x\left(\frac{\delta F}{\delta \gamma_{ij}}\frac{\delta G}{\delta\pi^{ij}}+\frac{\delta F}{\delta \phi_{A}}\frac{\delta G}{\delta\pi^{A}} - (F\leftrightarrow G)\right).\label{eq:DB_bi}
\end{equation}
For example, Hamiltonian fits in this category after excluding variables $u,\pi_u$, $u^i,\pi_{u^i}$ which can be done with the help of equations (\ref{eq:piu}), (\ref{eq:last}). Just opposite, we can express Hamiltonian by means of the potential $\tilde U$ and its derivatives with respect to variables $u,u^i$:
\begin{equation}
{\rm H}=\int d^3x\left( N\left(\tilde U -u\frac{\partial\tilde U}{\partial u}-u^i\frac{\partial\tilde U}{\partial u^i} \right)-N^i\frac{\partial \tilde U}{\partial u^i}\right),
\end{equation}
in this case more complicated formulas for Dirac brackets which take into account  variables $u,u^i$ are to be used, see Appendix \ref{S:DB}.

Hamiltonian of bimetric theory is not a linear combination of first class constraints, in contrast to bigravity and GR Hamiltonians, so we can use it for construction of conserved quantities. For example, if the background metric is flat, then the Hamiltonian density may be interpreted  as a density of energy, momentum or angular momentum, and all these quantities are ultralocal  functions of the corresponding lapse and shift $N,N^i$, two induced metrics $\eta_{ij},\gamma_{ij}$ and $u,u^i$.
 
\section{Conclusion}
In this article we have studied the theory of bigravity with a potential of the general form and built the Hamiltonian formalism, avoiding unnecessary noncovariance in defining the lapse and shift functions, which is an atavism of the notation in the ADM approach.  The potential in this work should fulfill only two essential conditions: the Hesse matrix $\mathbf{L}$ (\ref{eq:L}) is to be nondegenerate and 
equations (\ref{eq:pot1}), (\ref{eq:pot2}) are to be satisfied. Then constraints (\ref{eq:R}) and (\ref{eq:Ri}) are first class and provide us with the algebra of hypersurface deformations in Dirac brackets. 
The gravitation in this model has eight degrees of freedom, which together with the sign indefiniteness of the Hamiltonian indicates the presence of the Boulware-Deser ghost~\cite{BD}. The special potential choice proposed by de Rham-Gabadadze,Tolley~\cite{dRGT} is treated in a companion article~\cite{SolTchi2}.  

To the best of our knowledge, the Kucha\u{r} formalism has not been used previously in bigravity. The ADM analysis of bigravity\footnote{The first variant of a gravitational theory with two dynamical metrics appeared in the early 1970s~\cite{ISS}, but stayed unpopular.} was first proposed in~\cite{Kogan}, where it was also shown that this formalism is a generalization of the corresponding formalism of the massive gravity considered previously by Boulware and Deser~\cite{BD}. In the latter case, we have four second class constraints, which can be solved for the lapse and shift functions, and we can therefore eliminate these functions and stay with six pairs of canonical variables free of constraints. It was shown in~\cite{Kogan} that four new additional constraints appear in bigravity, and based on the theory invariance under general coordinate transformations, it was claimed that these constraints would be first class, but no calculations were done to support this claim.  Here we filled this gap, simultaneously taking a step toward a better understanding of bigravity with a special type potential~\cite{dRGT}, \cite{HR}, for which we cannot determine all four functions $u$, $u^i$ from four second class constraint equations (\ref{eq:last}).

{\it Note added.} Just after submission of this work to the journal, a preprint by Kluson appeared in arxiv.org~\cite{Kluson}, in which an analogous problem was considered independently and the same results were obtained. The main difference is that Kluson considered two important examples of the metric interaction potential but not the general case. Kluson later wrote another paper~\cite{Kluson2}, in which he confirmed his results for a more general potential  expressed in terms of the complete set of invariants of the matrix $g^{\alpha\mu}f_{\mu\beta}$. The de Rham-Gabadadze-Tolley potential was also discussed there. We give our conclusions concerning this case in the companion paper~\cite{SolTchi2}  where we also compare the different appoaches and results.  

{\bf Acknowledgements} One of the authors (V.O.S.) is grateful to the organizers and participants of Workshop on Infrared Modifications of Gravity  (ICTP, Trieste, 26 - 30 September, 2011) for stimulating atmosphere and to Prof. S. Randjbar-Daemi for hospitality during his visit to ICTP.

\newpage
\appendix
\section{First class constraints algebra in bigravity}\label{S:FC}
In the theory with the two dynamical metrics, as in GR, there are four first class constraints and their algebra is the algebra of hypersurface deformations. In computation of Poisson brackets between constraints  ${\cal R}$, ${\cal R}_i$ we treat
$u,u^i$
not as canonical variables, but as functions, because their conjugate momenta $\pi_u,\pi_{u^i}$ do not appear in the computation. Then potential $\tilde U$ in these calculations has nonzero Poisson brackets with gravitational momenta $\Pi^{ij},\pi^{ij}$ only, giving as results its derivatives with respect to induced metrics $\partial\tilde U/\partial\eta_{ij}$ and  $\partial\tilde U/\partial\gamma_{ij}$. First, consider equation (\ref{eq:bialg2}), generators ${\cal H}_i$ and $\bar{\cal H}_i$ are mutually commuting, and each of them separately satisfies equation (\ref{eq:alg2}), therefore, equation (\ref{eq:bialg2}) is also fulfilled. L.h.s. of equation (\ref{eq:bialg1}) can be given as follows
\begin{eqnarray}
\{{\cal R}(x),{\cal R}(y)\}&=&\{{\cal H}(x),{\cal H}(y)\}+\\
&+&\{u\bar{\cal H}(x),u\bar{\cal H}(y)\}+\{u^i\bar{\cal H}_i(x),u^j\bar{\cal H}_j(y)\}+\\
&+&\{u\bar{\cal H}(x),u^i\bar{\cal H}_i(y)\}+\{u^i\bar{\cal H}_i(x),u\bar{\cal H}(y)\}+\\
&+&\{{\cal H}(x),\tilde{U}(y)\}+\{\tilde{U}(x),{\cal H}(y)\}+\\
&+&\{u\bar{\cal H}(x),\tilde{U}(y)\}+\{\tilde{U}(x),u\bar{\cal H}(y)\}+\\
&+&\{u^i\bar{\cal H}_i(x),\tilde{U}(y)\}+\{\tilde{U}(x),u^i\bar{\cal H}_i(y)\}.
\end{eqnarray}
Lines from 1 to 3 can be easily computed by using formulas  (\ref{eq:alg1}), (\ref{eq:alg2}), (\ref{eq:alg3}): 
\begin{eqnarray}
 &=&\left(
 \eta^{ik}(x){{\cal H}}_k(x)+ \eta^{ik}(y){{\cal H}}_k(y)
\right)
 \delta_{,i}(x,y)+\\
 &+&u(x)u(y)\left(
 \gamma^{ik}(x){\bar{\cal H}}_k(x)+ \gamma^{ik}(y){\bar{\cal H}}_k(y)
 \right) \delta_{,i}(x,y)+\\
&+&u^i(x)u^k(y)\left({\bar{\cal H}}_i(y)\delta_{,k}(x,y)+ {\bar{\cal H}}_k(x)\delta_{,i}(x,y)\right)-\\
&-&u^i(x)u(y){\bar{\cal H}}(x)\delta_{,i}(x,y)+
u^i(y)u(x){\bar{\cal H}}(y)\delta_{,i}(y,x),
\end{eqnarray}
line 4 and line 5 give zero results, as they are antisymmetric in $x,y$ whereas each one contain $\delta$-function. Last line gives a contribution
\begin{equation}
2u^j\gamma_{jk}\frac{\partial\tilde U}{\partial\gamma_{ik}}(x)\delta_{,i}(x,y)-2u^j\gamma_{jk}\frac{\partial\tilde U}{\partial\gamma_{ik}}(y)\delta_{,i}(y,x).
 \end{equation} 
Taking into account relations 
 \begin{equation}
 f(y)\delta_{,i}(x,y)=f(x)\delta_{,i}(x,y)+f_{,i}\delta(x,y),\qquad \delta_{,i}(x,y)=-\delta_{,i}(y,x),\label{eq:delta}
 \end{equation}
 we can transform the result to the following form: 
\begin{equation}
\{{\cal R}(x),{\cal R}(y)\}=\left[Q^i(x)+Q^i(y) \right]\delta_{,i}(x,y),
 \end{equation}
 where
  \begin{equation}
  Q^i= \eta^{ik}{{\cal H}}_k-uu^i\bar{\cal H}+(\gamma^{ik}u^2+u^iu^k)\bar{\cal H}_k+2u^j\gamma_{jk}\frac{\partial\tilde U}{\partial\gamma_{ik}},
  \end{equation} 
  To fulfill equation (\ref{eq:bialg1}) it is necessary to have
 \begin{equation}
  Q^i= \eta^{ik}({{\cal H}}_k+{\bar{\cal H}}_k).
 \end{equation}
 As equation (\ref{eq:bialg1}) is to be valid for Dirac brackets, we can use second class constraints in the expression obtained for  $Q^i$, i.e.  replace $\bar{\cal H}$, $\bar{\cal H}_i$ by derivatives of the potential with respect for $u,u^i$, so this requires to treat equation (\ref{eq:pot2}) as a necessary condition for (\ref{eq:bialg1}).
  
 At last, let us check equation (\ref{eq:bialg3}):
\begin{eqnarray}
\{  {\cal R}_i(x),{\cal R}(y) \}
&=&\{
{\cal H}_i(x),{\cal H}(y)
\}+\\
&+&\{
\bar{\cal H}_i(x),u\bar{\cal H}(y)
\}
+\{
\bar{\cal H}_i(x),u^j\bar{\cal H}_j(y)\}+\\
&+&\{{\cal H}_i(x),\tilde{U}(y)\}+\{\bar{\cal H}_i(x),\tilde{U}(y)\}=\\
&=& 
{\cal H}(x)\delta_{,i}(x,y)+\\
&+& 
u(y){\bar{\cal H}}(x)\delta_{,i}(x,y)
+u^j(y)(\bar{\cal H}_i(y)+\bar{\cal H}_j(x))\delta_{,i}(x,y)+\\
&+& 2\left(\eta_{im}(x)\frac{\partial\tilde U}{\partial\eta_{mn}}(x)+\gamma_{im}(x)\frac{\partial\tilde U}{\partial\gamma_{mn}}(x) \right)\delta_{,n}(x,y)+\\
&+&\left[ 
\left(
2\eta_{im}\frac{\partial\tilde U}{\partial \eta_{mn}}+2\gamma_{im}\frac{\partial\tilde U}{\partial\gamma_{mn}} 
\right)_{,n}
-\eta_{mn,i}\frac{\partial\tilde U}{\partial\eta_{mn}}-\gamma_{mn,i}\frac{\partial\tilde U}{\partial\gamma_{mn}}
\right]
\delta(x,y).\nonumber
\end{eqnarray}
Proceeding in the same way to the previous calculation, and interpreting this bracket as Dirac bracket, we can use second class constraints in the expressions obtained. Then after exploiting formulas (\ref{eq:delta}), it is possible to demonstrate that in order to fulfill equation (\ref{eq:bialg3}) the potential has to satisfy condition (\ref{eq:pot1}).

\section{Second class constraints and Dirac brackets in bigravity}\label{S:DB}
To make our presentation compact let us introduce notations: $u^a=(u,u^i)$,  $\pi_a=(\pi_u,\pi_{u^i})$, $\bar{\cal H}_a=(\bar{\cal H},\bar{\cal H}_i)$, $a=1,..,4$.
We denote eight second class constraints as $\chi_A$,  $A=1,..,8$, so now we have
\begin{equation}
\chi_A=\left(\pi_a,\bar{\cal H}_a+\frac{\partial\tilde U}{\partial u^a}\right),
\end{equation}
Then matrix of Poisson brackets for these constraints has the following structure 
\begin{equation}
||\{\chi_A(x),\chi_B(y)\}||=
\left(
\begin{array}{cc}
\mathbf{0} & -\mathbf{L}(x)\delta(x,y)\\
\mathbf{L}(x)\delta(x,y) & \mathbf{K}(x,y)
\end{array}\right),
\end{equation}
where
\begin{equation}
 \mathbf{L}_{ab}(x)=
\frac{\partial^2\tilde U}{\partial u^a\partial u^b}(x) ,\label{eq:L}
\end{equation}
\begin{equation}
\mathbf{K}_{ab}(x,y)=\left\{
\bar{\cal H}_a(x)+\frac{\partial\tilde U}{\partial u^a}(x),  
\bar{\cal H}_b(y)+\frac{\partial\tilde U}{\partial u^b}(y)
\right\},
\end{equation}
we suppose that in the general case matrix $\mathbf{L}$ is invertable, then matrix $||\{\chi_A,\chi_B\}||$ is invertable too, and its inverse is as follows
\begin{equation}
\mathbf{C}
=\left(
\begin{array}{cc}
 \mathbf{L}^{-1}(x)\mathbf{K}(x,y)\mathbf{L}^{-1}(y) & \mathbf{L}^{-1}(x)\delta(x,y)  \\
 -\mathbf{L}^{-1}(x)\delta(x,y) & \mathbf{0}
\end{array}
\right).
\end{equation}
Dirac brackets are given in the following way
\begin{equation}
 \{F,G\}_D=\{F,G\}-\int dx\int dy\{F,\chi_A(x)\}\mathbf{C}^{AB}(x,y)\{\chi_B(y),G\}
\end{equation}
and in cases when both functionals $F$, $G$ do not depend on variables $u,u^i,\pi_u,\pi_{u^i}$, these brackets coincide with the Poisson ones. Therefore, if constraints are explicitely solved for $u,u^i$, and these variables are replaced by constraint solutions, then for the rest of variables $\eta_{ij},\Pi^{ij}$, $\gamma_{ij},\pi^{ij}$, $\psi_A,\Pi^A$, $\phi_A,\pi^A$ Dirac brackets coincide with canonical Poisson brackets:
\begin{equation}
\{F,G\}_D=\int d^3x\left(\frac{\delta F}{\delta \eta_{ij}}\frac{\delta G}{\delta\Pi^{ij}}+\frac{\delta F}{\delta \psi_{A}}\frac{\delta G}{\delta\Pi^{A}}+
\frac{\delta F}{\delta \gamma_{ij}}\frac{\delta G}{\delta\pi^{ij}}+\frac{\delta F}{\delta \phi_{A}}\frac{\delta G}{\delta\pi^{A}} - (F\leftrightarrow G)\right).\label{eq:DB1}
\end{equation}
But sometimes an explicit solving of constraints may be difficult, then one may write bigravity Hamiltonian by means of potential $\tilde U$, which is a function of variables $u,u^i$ and of the two induced metrics $\eta_{ij},\gamma_{ij}$ (\ref{eq:66}).
In such a situation one has to use nontrivial Dirac brackets in obtaining Hamiltonian equations or in solving other problems:
\begin{equation}
\{\gamma_{mn}(x),u^a(y)\}_D=-{\mathbf{L}^{-1}}^{ab}(y)\{\gamma_{mn}(x),\bar{\cal H}_b(y)\},
\end{equation}
\begin{equation}
\{\pi^{mn}(x),u^a(y)\}_D={\mathbf{L}^{-1}}^{ab}(y)\left(
\frac{\partial^2\tilde U}{\partial u^b\partial \gamma_{mn}}\delta(x,y)-
\{\pi^{mn}(x),\bar{\cal H}_b(y)\}\right),
\end{equation}
\begin{equation}
\{\Pi^{mn}(x),u^a(y)\}_D={\mathbf{L}^{-1}}^{ab}\frac{\partial^2\tilde U}{\partial u^b\partial\eta_{mn}}\delta(x,y).
\end{equation}

\end{document}